# Causes of high-temperature superconductivity in the hydrogen sulfide electron-phonon system

Nikolay Degtyarenko, Evgeny Mazur

*National Research Nuclear University (Moscow Engineering Physics Institute), Kashirskoe sh.31, Moscow 115409, Russia*

## 1. Introduction

The transition to a superconducting phase was observed in the metallic phase of hydrogen sulfide at pressure $P = 170$ GPa at $T = 190$ K in [1]. High $T_c$ value in hydrogen sulfide under pressure is a property exclusively related to the electron-phonon system. By means of ab-initio modeling we found the structure of $SH_2$, which gives a pressure change in the characteristics similar to the results presented in [1]. The primitive orthorhombic phase cell containing 1 sulfur atom and two hydrogen atoms was found. The cell symmetry appears to be I4/MMM. In fact, the stable phase was found in the numerical calculations by withdrawing the central sulfur atom and the corresponding hydrogen atoms from the cubic cell $SH_3$ structure [2]. Further, the geometry of the structure is optimized for the data values of the external pressure with the use of the principle of minimum total energy of the system. Optimization results are presented in Fig. 1. We are not aware of any studies concerning such a structure in the literature. In a large range of pulse values, the phonon spectrum structure with symmetry I4/MMM has no imaginary frequencies. The phase stability or the existence of the metastable phase is checked by a full calculation of the dispersion behavior of all branches of the phonon spectrum. There should not arise either any imaginary frequency in the phonon spectrum or areas with the pronounced softening of the phonon spectrum in a specific Brillouin zone (BZ) area in a stable or metastable phase. We used a method of density functional theory (DFT) with the plane-wave basis set and with Generalized-Gradient Approximation (GGA) of the exchange correlation

functional – PBE; for sulfur atoms, we used a pseudopotential that preserves the norm.

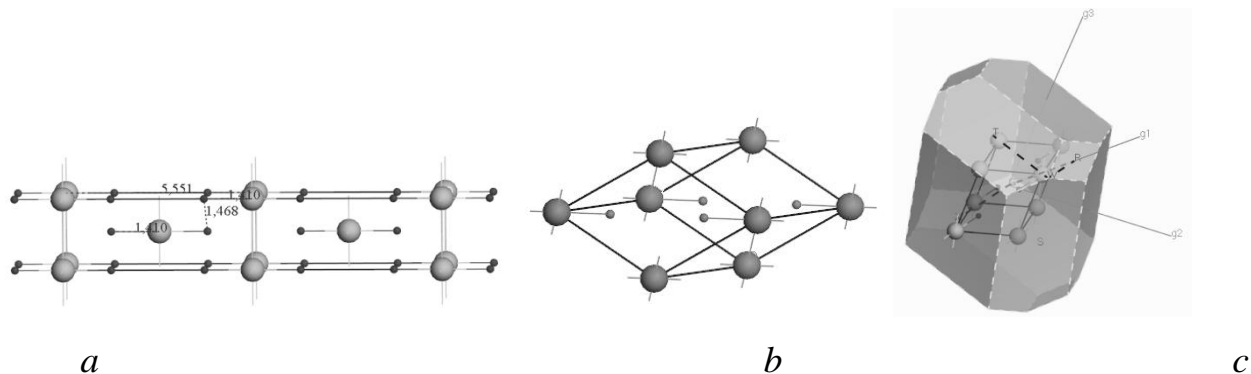

| a | b | c |

Fig.1.The SH$_2$ structure study. Sulfur atoms are represented as balls of a larger size; the results of calculations for the pressure $P$ = 170GPa; $a$ – the initial cubic cell; $b$ – the primitive cell with the symmetry I4/MMM (D4H), $a = b = c = 2,99$ Å; $\alpha = 64,08^o$, $\beta = 141,53^o$, $\gamma = 130,89^o$; c – Brillouin zone corresponding to the primitive cell

## 2. Results.

The results of ab-initio calculations of the structural, electron, and phonon characteristics of SH$_2$ in the pressure range of 100 – 225 GPa can be summarized as follows. The optimization of both the unit cell configuration and the internal basis with the principle of minimum total energy of the system at a predetermined pressure value at the boundaries of the system has been done for each value of the hydrostatic pressure. It was assumed that the system has a metallic character in all calculations. Graphs of the band structure and of the density of electron states (DOS) for two pressures $P$ = 165 GPa are presented in Fig. 2. The pictures for the pressure $P$ = 175 GPa are very similar to the pictures in Fig 2, with the exception of the Fermi level located at the local minimum of DOS.

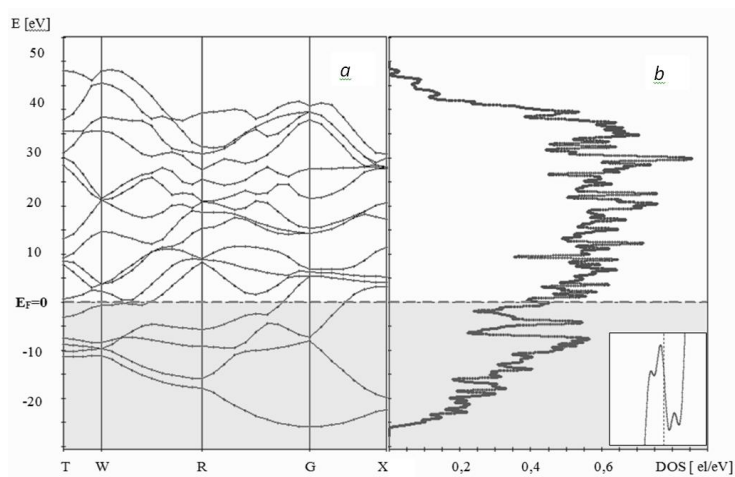

Fig.2. *a* – Electron band structure of $SH_2$ at the pressure $P = 165$ GPa, $E_F = 0$ indicates the Fermi level; *b* – the density of electron states for $SH_2$ at the pressure $P = 165$ GPa, $DOS(E_F) \approx 0.43$; the Fermi level is located on the slope of the DOS peak (see. box on the plot)

As we can see from Fig. 2 the Fermi level is located near the electron density peak of the *s*-type states in both cases corresponding to the high $T_c$ value. The majority of *s*-type states refer to hydrogen. In this case a sharp decrease in the density of electron states is distinctly seen above the Fermi energy in the system. The phonon dispersion and the density of phonons (DOS) are shown vs energy for the pressure $P = 165$ GPa at Fig.3, *a-b*. For $P = 175$ GPa the pictures of phonon dispersion and the density of phonons (DOS) are very similar to the pictures in Fig.3. As one can see from Fig.3 the hydrogen sulfide phonon spectrum in the orthorhombic phase under these pressures does not contain the areas similar to the phonon nodes. These results indicate the stability (or, at least, metastability) of the analyzed phase. Two narrow peaks with phonon mode energies can be seen at the phonon density of states in a high-frequency region. Only one of these two peaks is active in the infrared absorption.

The calculated phonon spectrum is shown in Fig. 4 for the pressure $P = 180$ GPa. As one can see, one of the acoustic branches of the phonon spectrum tends to a zero value for this pressure, and the calculated phonon frequency turns to "negative" values (imaginary values in reality), indicating the instability of the selected orthorhombic phase near the given pressure. This result is fully consistent with the experimental results [1], where the phase transition point for this system was detected at the pressure $P = 180$ GPa. The changes in the electron density distribution should be the result of structural changes of the unit cell with pressure in the test phase. The detailed analysis of changes of this structure with increasing pressure shows that the specific feature of this structure should be the formation of the system of parallel planes with the full-body concentration of hydrogen atoms in these planes at the pressure $P = 150 - 170$ GPa.

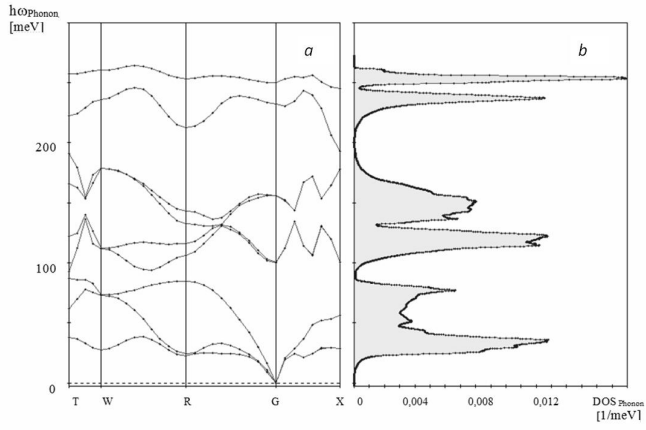

Fig.3. *a* – phonon dispersion relation for SH$_2$ at the pressure $P$ = 165 GPa; *b* – phonon density of states (DOS$_{Phonon}$) for SH$_2$ at the pressure $P$ = 165 GPa

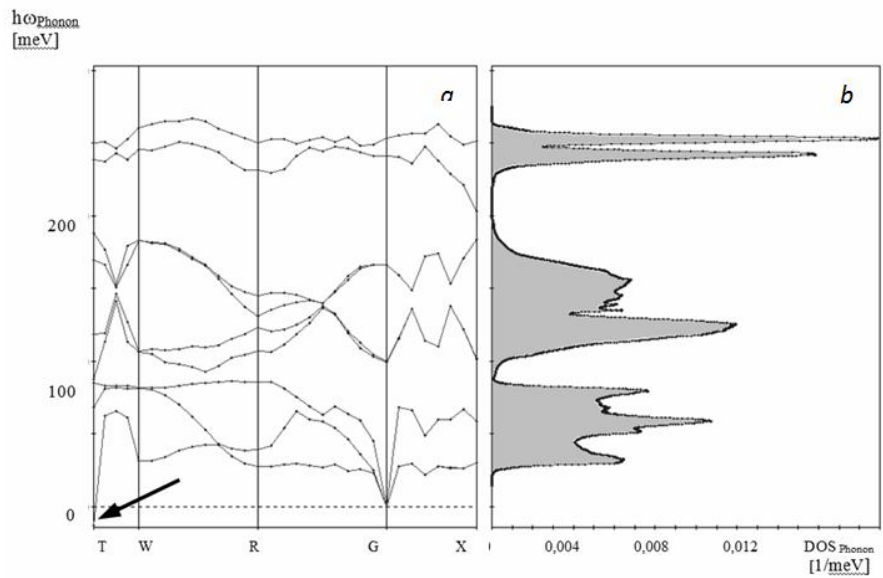

Fig.4. *a* – SH$_2$ phonon dispersion relation at the pressure $P$ = 180 GPa; the arrow indicates the acoustic branch of the phonon spectrum, turning to a negative value and thus indicating the instability of the selected phase at the pressure value in the vicinity of the given set points (negative values indicate imaginary frequency in the program); *b* – Phonon DOS, i.e. SH$_2$ phonon density of states at the pressure $P$ = 180 GPa

As a result, the electron properties of the system acquire a quasi-two-dimensional character. Fig. 5 *a-b* shows the pictures illustrating the nature of the vibrations of H atoms in the unit cell for pressure $P$ = 170 GPa in the lattice for this stable phase. Only three modes have non-zero intensity of interaction with infrared light from the six lattice vibrational modes (ν: 309.43; 309.43; 1169.91; 1169.91; 2306.72; 2420.94 [cm$^{-1}$]). Two modes with maximum frequencies correspond to the conditions for the high critical temperature $T_c$ for the given SH$_2$ phase. The vibrations of atomic plane consisting of hydrogen atoms with the frequency ν$_1$ =

2420.94 [cm$^{-1}$] have zero intensity in infrared radiation. These oscillations are in antiphase (Fig. 5$a$). In -phase oscillations of atomic plane consisting of hydrogen atoms have the frequency $\nu_2 = 2306.72$ [cm$^{-1}$] (Fig. 5$b$) with a maximum intensity in infrared region.

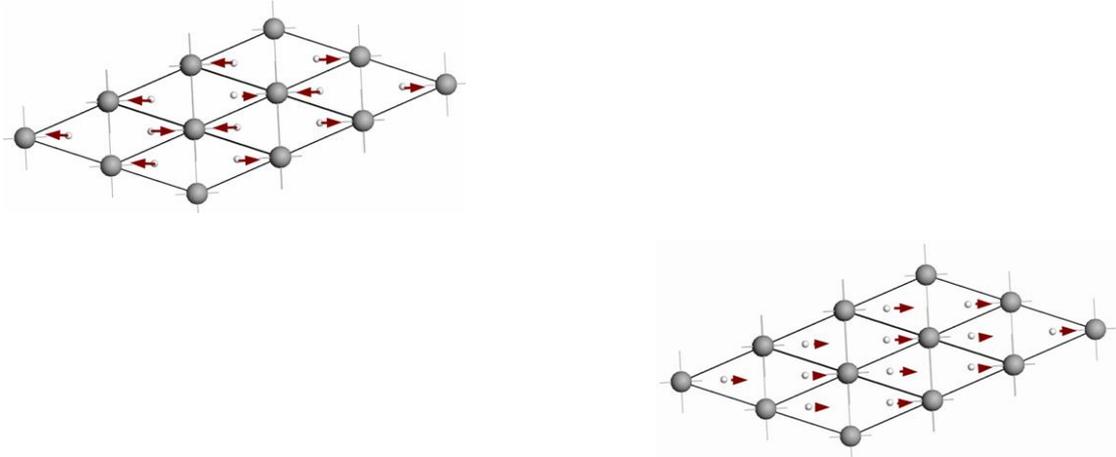

Fig.5 $a$ – The picture of the directions of displacements of hydrogen atoms (small balls) in the SH$_2$ structure at the pressure $P = 170$GPa with the frequency $\nu_1 = 2420.94$ [cm$^{-1}$]. These oscillations are in antiphase, $b$ – The picture of the in-phase displacements of the hydrogen atoms in the SH$_2$ structure at the pressure $P = 170$ GPa with the frequency $\nu_2 = 2306.72$ [cm$^{-1}$] (the highest peak intensity in IR).

## 3. Conclusions

- The compressed $SH_2$ crystallite with orthorhombic elementary cell having symmetry I4/MMM is considered. We found that a sulfur atom and two hydrogen atoms are positioned on a single line within the same cell. H atoms are located on both sides of the two sulfur atoms at the distance of 1.4 Å. Such a structure is formed at the pressure exceeding 100 GPa in the cubic $SH_2$ phase.

- DOS and phonon dispersion relations contain imaginary frequencies when pressure is approximately equal to 100 GPa.

- None of the phonon branches in $SH_2$ structure with elemental orthorhombic cell having symmetry I4/MMM vanishes at any point of the Brillouin zone in the pressure region $P = 125$–$175$ GPa.

- At the pressure P=180 GPa one of the acoustic branches of the phonon spectrum vanishes at the point T in Brillouin zone, indicating the phase transition point of the system. So, the metastability of the investigated orthorhombic structure $SH_2$ with increasing pressure in the region P = 125–175 GPa is shown to emerge.

- With pressure increasing, the hydrogen atoms belonging to the sulfur atoms of different neighboring cells move in the collinear directions reducing distance among them. When the pressure approaches $P \sim 170$ GPa, a single plane is formed consisting of hydrogen atoms only.

- If the pressure exceeds the optimal values, the counter collinear motion destroys the above mentioned plane during the unit cell deformation. The plane is divided into two planes. H atoms spread apart, but the electron density between them continues to grow due to the compression of the unit cell.

- In the specified pressure range, the electron PDOS changes at the Fermi level. In the work it is meant as complete electron and partial density of states over the atoms and over the $s$- and $p$- electrons. But one cannot argue that the Fermi level is located at the position of the maximum peak within the density of states.

- At the pressure $P = 180$ GPa there arises some softening of the phonon modes, and a specific feature on the phonon dispersion curves in the symmetry point T of the inverse space is calculated that may indicate an initial stage of structural instability.